# The Wide-field Spectroscopic Telescope: optical designs for the low-resolution multi-object spectrographs

Will Saunders[1*], Kjetil Dohlen[2], Dimitri Buffat[2], Roelof S. de Jong[3], Olaf Iwert[4], David Lee[5], Andrea Bianco[6], Ross Zhelem[1], Etienne Burtin[7], Virgile Meyer[7]

[1]Astralis, Australian Astronomical Optics, Macquarie University (Australia). [2]Aix Marseille Univ, CNRS, CNES, LAM, Marseille (France). [3]Leibniz-Institut für Astrophysik Potsdam (Germany). [4]European Southern Observatory (Germany). [5]UK Astronomy Technology Centre (United Kingdom). [6]INAF (Italy).[7]Univ. Paris-Saclay, CEA-IRFU (France).*will.saunders@mq.edu.au

## ABSTRACT

In MOS Low-Resolution mode, WST has a 'spectral etendue' per fiber (etendue x spectral resolution elements) twice as large as any existing MOS instrument, and a total spectral etendue ~20 times larger. Combined, these present enormous challenges for the spectrograph design. Initial designs were based on (a) a novel F/0.775 Folded Solid Schmidt camera with 6cm detectors, (b) an F/1.15 dioptric camera with 6cm detectors, and (c) an F/1.31 dioptric camera with 9cm detectors, all with YAG field-flatteners. However, the imperative of fast camera speed has greatly diminished, as the projected cost and read-noise of large format CMOS detectors is projected to decrease greatly. The adopted baseline solution has 4 dioptric F/1.64 cameras, each with two doublets, a YAG field lens and 9cm x 9cm detector. Theoretical image quality is 8 microns rms radius. The design seems readily athermalised, potentially allowing passive temperature control. Each spectrograph accepts 500+ fibers. Expected RoM costs (scaled from earlier designs) suggest a total cost <4M Euro per spectrograph.

## 1. INTRODUCTION

WST MOS-LR will have ~30,000 fibers, an order-of-magniude increase over current systems. But also, each fiber will have ~4000 resolution elements, fed by a 12m telescope in natural seeing. This results in a 'spectral etendue' per fiber almost a factor 2 larger than any current multifiber system, while the total spectral etendue is over 20 x larger. These features create unique challenges for the spectrograph design – the design is technically very challenging, but nonetheless, needs to be simple, affordable, compact, and readily installed, aligned and maintained.

**Design Drivers**

The required wavelength coverage is 370-930nm (originally 370-980nm), with a required resolution $R > 3000$ everywhere. In principle, 2760 resolution elements should suffice. However, this number is increased for various reasons:

- Within each spectrograph channel, the resolution is effectively constant in nm, so $R$ is higher than the requirement everywhere except right at the blue end.
- A higher resolution (say $R \sim 5000$) is desirable in the far red, both for minimising *OH* contamination, and for achieving ~1km/s stellar radial velocities from the calcium triplet. The impact of resolution on far red performance is considered further in the appendix.
- Extra resolution elements are required to allow for dichroic crossover widths of 2-3% (depending on AOI).
- Pin-cushion distortion and spectral curvature mean that the detectors cannot be fully utilised.

In practice 4000+ spectral resolution elements are desirable. And since faster camera speeds increase the number of fibers that can be accommodated, there is a natural tendency in any case to push any design as fast as possible.

Available detector sizes are assumed to be 61mm x 61mm ('6cm') or 92mm x 92mm ('9cm'), both with assumed pixel sizes of 15um. For CCDs, these numbers have remained stable for many years, and are hence likely to be to be relevant for CMOS detectors also.

Given these numbers, and feasible camera speeds, then there are only 4 realistic options. Each option has been considered, and the name of the resulting design is in brackets. The spectral etendue per spectrograph of each option is conveniently measured in 'DESIspecs', the number of arms x the detector area x the square of the camera speed, vs the same quantities for DESI.
A 3-armed design with 6cm detectors and F/~0.8 cameras (**FSS,** 5 DESIspecs);
A 4-armed design with 6cm detectors and F/~1.1 cameras (**4x6,** 3 DESIspecs);
A 3-armed design with 9cm detectors and F/~1.2 cameras (**3x9,** 4 DESIspecs);
A 4-armed design with 9cm detectors and F/~1.65 cameras (**4x9,** 3.4 DESIspecs).

Initially, the first three options were pursued, on the basis that detector costs and read-noise drove the design to the fastest speeds possible. However, the cost difference of 9cm versus 6cm detectors has greatly diminished, and projected read-noises have declined greatly. Also, 9cm CMOS detectors have been selected for WST IFS, and so form a desirable choice for MOS-LR also. The 4x9 design was selected at a Trade Study in March 2026. and forms the current baseline design. All four designs are described below.

## 2. DISPERSER CONSIDERATIONS

Disperser options are considered in detail in a separate paper [1]. However, they are relevant to the optical design process itself, and so are briefly considered here. For VPH gratings (as assumed at the outset of this study), both the peak efficiency and the angular bandwidth are remarkably constant for grating angles in the range of interest (20-30º). This means that for fixed beam-size and overall wavelength coverage, the number of arms, camera speed and detector sizes have little effect on VPH efficiency. This is illustrated in Figure 1, which shows VPH efficiency curves supplied by Wasatch Photonics for 3 and 4-armed variants with the same beam size, showing only modest differences in overall throughput.

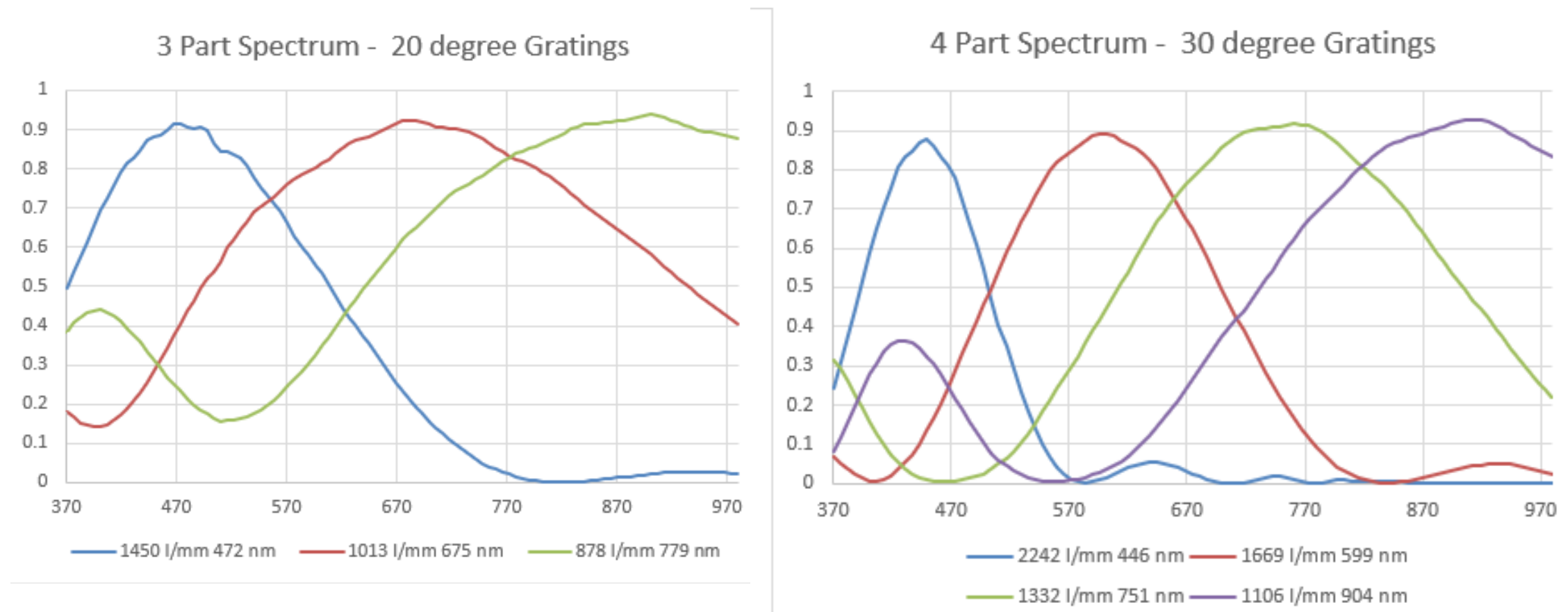

**Figure 1: Preliminary VPH grating designs from Wasatch Photonics, for 3 and 4 channels with fixed beam size. The efficiency bandwidth and dispersion mostly cancel, to give remarkably constant average and peak throughputs. However, better *s-p* phasing means that lower dispersion gives slightly better overall efficiency in this regime.**

However, the beam-size is directly relevant to average VPH efficiency, because a larger beam implies smaller field angles and hence increased angular bandwidth. This means that in principal, catadioptric designs have better grating efficiency for a fixed number of arms and spectral coverage, partially compensating for their inescapable obstruction losses. In practice, beam size is determined almost entirely by image quality, optics availability and cost considerations, so this degree of freedom is not as useful as it might seem.

## 3. COMMON FEATURES

Telescope speed is F/3.395 with an assumed 12m pupil. The fiber diameter is assumed to be 1", or 195um. The wavelength coverage is 370nm-980nm, except for 4x9 where it is 370nm-930nm (the requirement was changed in March 2026). All designs share many common features, as follows.

**Collimator considerations**

The optimal collimator speed is not obvious, and is still subject to revision. The WST telescope mirror is roughly hexagonal, and the beam entering the fibers has off-axis vignetting, at M2 and L1, to preserve 'effective pupil-centricity' [2]. Figure 2 shows the azimuthally summed far-field light profiles entering and leaving the fibers, using a diffusive FRD model giving the same overall FRD performance as 4MOST. The nominal collimator over-filling losses are ~8% at F/3.15 and ~10% at F/3.25, a small difference considering the 6% difference in etendue.

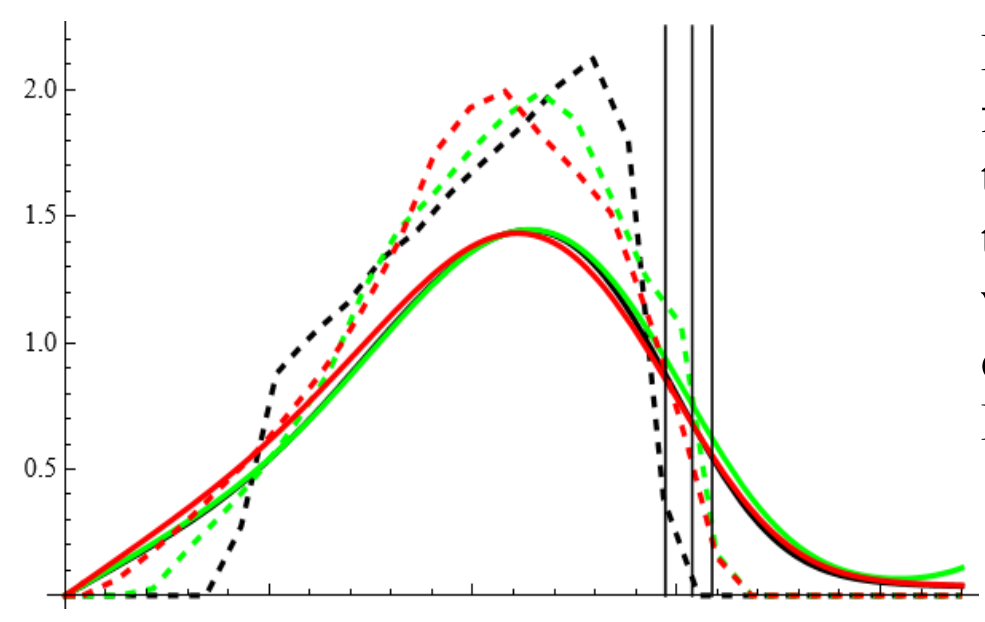

**Figure 2. Azimuthally summed far-field light vs NA profiles for WST, at field angles 0, 0.7°, 1° (black, green, red), for the telescope itself (dashed lines), and as estimated leaving the fibers after Focal Ratio Degradation (solid lines). The vertical lines from left to right represent the telescope speed of F/3.395, the dioptric collimator speed of F/3.25, and the FSS collimator speed of F/3.15**

The collimator is always an off-axis Schmidt, with the slit unit on one side, and all the cameras on the other side. This is similar to that used in 4MOST LRS [3]. It gives better throughput than an on-axis Schnidt, a long collimated space for dichroic placement, and simplifies alignment, focus and maintenance.

The slit is curved axially, and also has lateral 'smile' with 5-10mm sag. The slit is gelled to a slit lens, allowing freedom in pupil placement and reducing air/glass losses. The collimator mirror is spherical. There is very small clearance between the inner edge of the slit lens and the return beam, but this still allows a simple flag shutter, as used in AAOmega [4], and could be increased as needed.

There are two or three dichroics, with substrates of fused silica.

The correctors are freeform, with ~7mrad max aspheric slope. The sides are trimmed, increasing clearances and making the CoC of the collimator mirror accessible for collimator alignment.

All designs have the disperser incorporated into a one-sided grism, saving 2 air/glass surfaces. The input and output grating angles are set to be equal, but with one in air and one in glass. This implies non-Littrow use (and hence no Littrow ghost [5]), but does not introduce anamorphism, so the camera lenses remain fully utilised.

The cameras always have a field-flattener lens, which also forms the dewar window, and this is always of Yttrium Aluminium Garnet (YAG). YAG seems an ideal material for both purposes - it has high index ($n_D$=1.83) with low dispersion ($v_D$=53), it is transparent and non-birefringrent; it is non-radioactive, 30% stronger than fused silica, and 9 times more thermally conductive. Blanks of the requisite size are available from LasetTec (China).

The layout on the detector always puts the red end of the bluest channel along the edge of the detector, and the blue end at the ends of the slit in the opposite corners. Since the spectral curvature is weakest for this channel, but stronger than the pin-cushion distortion, this ensures that all spectra are entirely on the detectors for all channels.

All designs use flat detectors. Curved detectors (both spheric and cylindiric) were tried in the optical design, several times. Cylindrical curvature was not found to be useful, because the slit is of necessity pupil-centric for fiber-fed systems. Spherical curvature brings only a modest benefit at viable curvatures (RoC/diagonal > 3.45), though higher curvatures, with RoC ~ camera focal length, would be transformative [6].

All designs were laid out to have the equal resolution $R$ in each arm except the reddest, and the same resolution in nm for the two reddest arms, or as close to this as the requirement $R > 3000$ and feasible camera speed allowed.

All quoted image qualities asume that the beam coming out of the fiber is apodised, in a manner crudely consistent with the 4MOST FRD data.

## 4. FOLDED SOLID SCHMIDT

### Overview

The first option has 3 arms with 6cm detectors and ~F/0.8 cameras. This is faster even than the catadioptric cameras used for PFS and MOONS (F/1.09 and F/0.95 respectively). Also, classic Schmidt-type cameras are not attractive for a massively multiplexed system like WST, because the entire camera must act as a dewar (with huge cost and operational implications), unless the dewar sits in the beam, which then increases the already significant obstruction losses.

Faster speeds can be achieved through the use of a solid Schmidt camera. This offers an $n^2$ improvement in speed over a classic Schmidt ($n$ is the camera body refractive index) [7]. Detector access can be improved by folding the design, (as sometimes done for classic Schmidts [8]), giving access from behind and to one side of the detector. This results in a Folded Solid Schmidt (FSS) camera, which to our knowledge has not been proposed previously. The camera body forms half a solid cube, with an input surface, a fold mirror, a main mirror, and a field-flattening lens which also acts as dewar window. There are two inactive side surfaces, which allow for camera and dewar mounting. Fused silica is the only feasible material for UV use, and is assumed for all arms.

There is no obvious way to correct for chromatic aberrations within the camera body itself. The adopted solution is to mount the disperser in a grism, which is bonded to a flint/crown doublet. This corrects the chromatic aberrations while also providing some of the camera power[1].

The convex field-flattening lens must have higher index than fused silica[2], as well as having suitable properties as a dewar window, and YAG seems by far the best option. This lens must be optically coupled to the camera body by an oil or gel (to avoid TIR); an oil-coupling is awkward to seal, hence gel is much preferred.

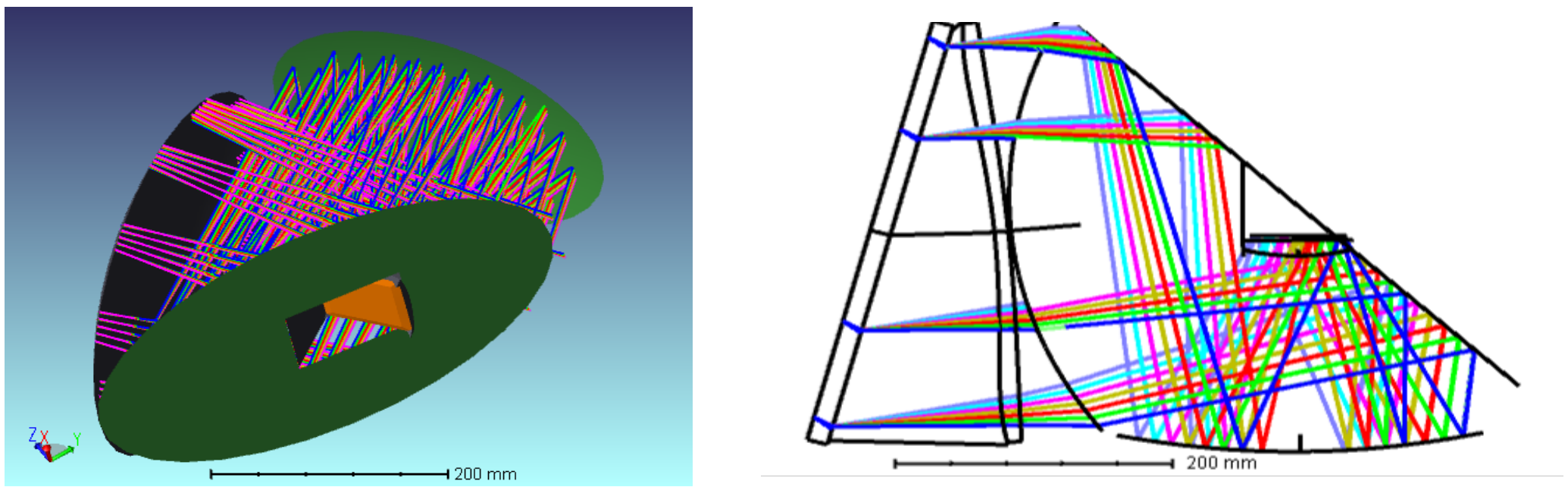

**Figure 3. Layout for the FSS camera. There is a powered entrance window, a fold mirror, a main mirror, a YAG field-flattening lens in a recess, and the detector.**

---

1 This is needed at low dispersion. At high dispersion, the reduced wavelength range in each channel makes a doublet unnecessary

2 A concave air space is possible in principle, but does not work at all, (image quality and TIR); and a dewar window is still required in any case.

The speed is limited by geometrical considerations, not image quality. A speed of F/0.775 allows plausible clearances, while exceeding the resolution requirements. This gives each spectrograph 5 x the etendue of a DESI spectrograph.

**FSS details**

Collimator speed is F/3.15, beam-size is set to 300mm, slit-length is 245mm, camera speed is F/0.775. The collimator and dichroic layout is as described in Section 3.

The gratings are mounted on a flat substrate of S-FSL5Y. The superstrate is a wedged plano-convex lens of S-FSL5Y, aspheric on the powered surface. This is bonded to a PBL35Y lens, aspheric on both surfaces. The asphericity on the internal face is very large (aspheric slope up to 250mrad!), but the required surface precision and smoothness are very low, because of the index-matched glue[3]. The CTEs are compatible. The external asphere is much milder, aspheric slope < 50mrad.

There is then an air-gap to the fused silica camera body. This consists of an entry surface, a fold mirror, a main mirror, and a partially recessed exit surface. The entry surface is coaxial with the grism lenses and is mildly aspheric (<18mrad). The fold mirror is shown here at 43° about the spatial-axis, which maximises the clearance between the input surface and the reflected beam. This surface would likely be dialectrically coated, which should give a better reflectivity than a metal coating over the restricted wavelength range in each camera. The next surface is the main camera mirror, mildly aspheric (< 13 mrad). Again, it is assumed to be dialectrically coated. The final surface of the camera body is the powered and very strongly aspheric (up to 350mrad!) seat for the dewar window. Polishing this surface is complicated by it being recessed into the camera body. However, this surface will have coupling gel between it and the dewar window, with almost perfectly matched index, so the tolerance on both figure and roughness are extremely low, and it is proposed that it be simply laser-assisted diamond cut.

The final optic is the YAG field-flattening lens, which also acts as dewar window. The lens is bi-convex and bi-aspheric, with slope < 67mrad on the 2nd surface. Efforts to make the 2nd surface flat (or even spherical or conic) were unsuccessful. The YAG field-flattener is gelled into the seat in the camera body. The gel must be thick enough to allow a focus tweak of some tens of microns during set-up (to correct for any residual differences in focus/tip/tilt between cameras that cannot be corrected at the slit).

The lens is cut to 80mm square, with rounded corners, to minimise vignetting. This gives a 5mm mounting allowance around the optical aperture. The central thickness is 12-15mm, and corner thickness ~2.5mm. The thickness at the centers of the sides of the optical aperture (which take most of the load) is >~7mm. The design is tolerant to increasing these thicknesses, the issue is the increased lens size and hence increased vignetting.

The lens must be correctly positioned on the dewar, to within ~50um. The proposed solution is to mill the outer ~5mm of the lens surface flat, to a depth of ~1mm; the square aperture of the dewar is made the same size (~70mm). Then the lens sits in this aperture, and this also gives a flat surface for vacuum sealing, and increases the available clearance to the detector. The O-ring sits in a trough, and must be sized so that the lens sits in full contact with the dewar mounting surface when under vacuum load.

There is then a 2-3mm central gap to the detector surface. The detector needs to be tilted about the spatial axis, by up to 0.5°. The smallest clearance at the edge of the detector is 2.15mm. The design is quite tolerant to this gap, the issue is once again the vignetting.

**Image quality, resolution, throughput**

The image quality is remarkably good, with rms radius < 5um almost everywhere (Figure 4). The projected fiber diameter $d$ is 49um. To this must be added as-built errors, detector flatness, and charge diffusion.

The resolution meets the requirements of $R > 3000$, and gives $R > 5000$ at $\lambda > 850$nm.

The throughput is good, by catadioptric standards. The losses come from:
- 7 air/glass surfaces, plus 1 or 2 dichroics;

3 As done successfully for AAOmega [5]

- ~130mm of *i*-line glass;
- Three mirrors, two of them immersed and dialelectrically coated.
- Vignetting by the detector package, but this is partially in the shadow of the top-end obstruction (Figure 6).
- The VPH efficiency is good, essentially because the beam is so large (300mm), reducing the grating angles to (20°/20°/25°), versus the modest channel bandwidths $\lambda_{max}/\lambda_{min}$ ~ 1.45/1.45/1.32.

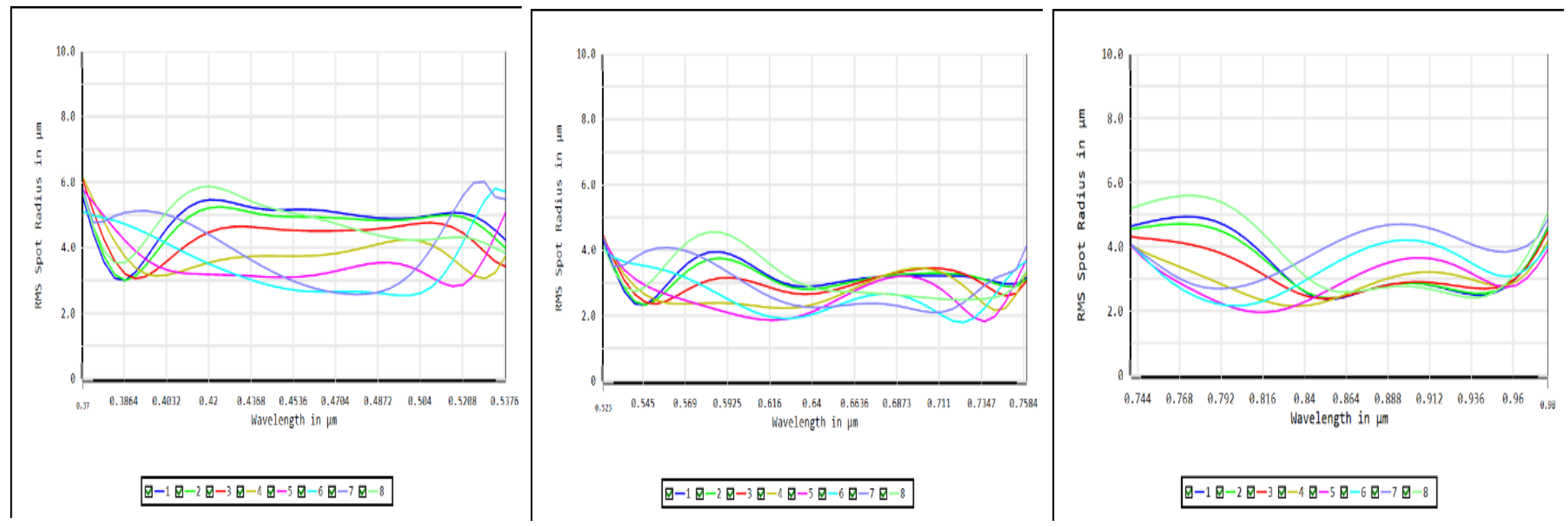

**Figure 4: RMS radius vs color and slit position for the 3 arms, for the FSS design. Projected fiber**

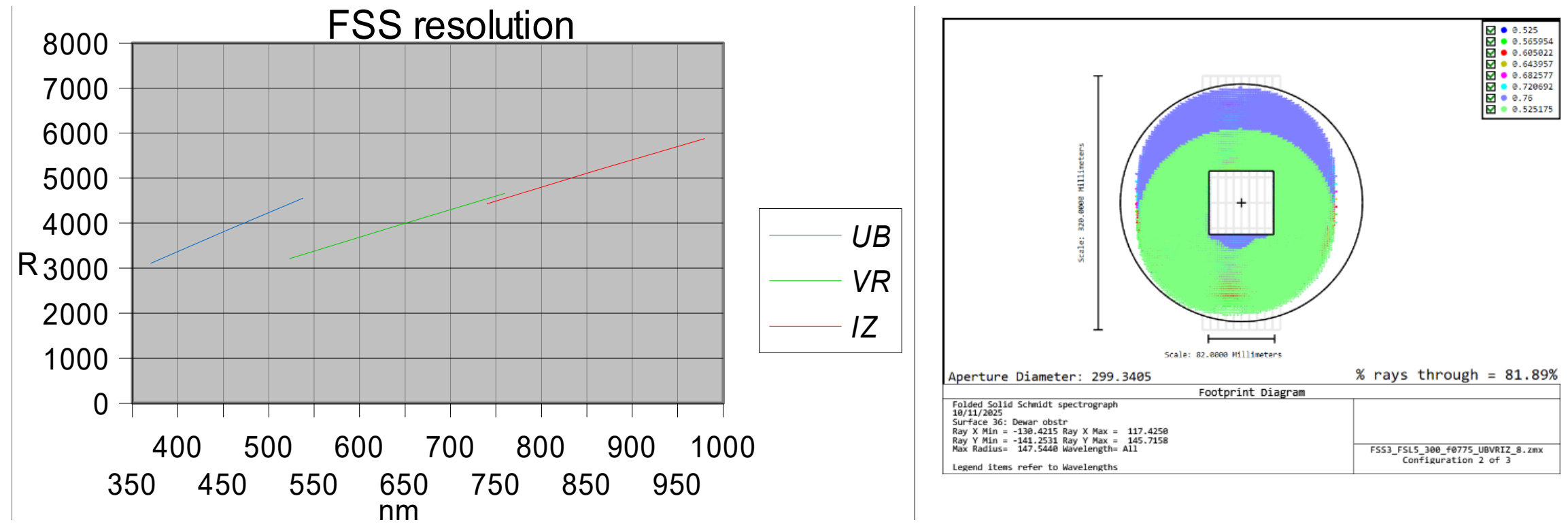

**Figure 5.(left) FSS spectral resolution *R* vs wavelength**

**Figure 6: (right) Beams as they pass the detector obstruction within the FSS camera. The nominal obstruction due to the telescope top end (on-axis and in the absence of FRD) is shown. For most field angles and slit positions, the shadow of the top end obstruction is entirely within the detector obstruction, significantly reducing their combined effect. The additional detector vignetting depends on wavelength and slit position, but is always <~10%.**

### Thermal issues

A potential concern for the FSS design is whether the fused silica camera body will have large thermal gradients due to the radiative losses to the dewar, and that the large *dn/dT* of fused silica could then compromise the optical performance. A back-of-envelope calculation suggests that this should not be a significant issue. YAG has a high thermal conductivity (12.9W/mK), and the field flattener is in good thermal contact with the dewar body via its mounting, and with the camera body via the coupling gel. The heat load into the detector is of order 2W, giving an expected maximum temperature difference between center and edge of lens of ~-4K. The maximum expected path difference for beams passing above the field lens is then <¬3um, small compared with eg the detector flatness. A more detailed thermal study, confirming this result, is presented separately [9].

**Alignment and focus**

The alignment of fast spectrographs is notoriously difficult, with typical tolerances of 10-20um. And in this case, we will have large numbers (40+) of spectrographs, each with 3 cameras. However, the $n^2$ improvement in speed offered by the solid Schmidt design applies also to the camera tolerances. While a full tolerance analysis has not yet been done, some simple sanity checks have been carried out. The most difficult lateral alignments are within the camera, and amount to ~20um.

However, it seems implausible to manufacture and align the dewar to the required <10um precision required for detector focus. The design offers no convenient focussing mechanism, other than to move the detector within the dewar (and piezo-electric focussing mechanisms are available for cryogenic environments [10]). If this is deemed too unreliable, it is proposed that the entire dewar and YAG lens be adjustable in piston, tip and tilt, albeit with a very small throw (say 25um max), compatible with a ~100um coupling gel thickness, allowing individual camera focus at this level.

**Manufacture and cost**

If it turns out not to be practicable to diamond-cut the recessed field-flattener lens seat, it would be possible to make the camera body in three or four pieces as shown in Figure 7 (input surface, main mirror, lower body, upper body); this allows the lens seat to be polished classically. The fold mirror can then be post-polished, after the components are assembled. A nice feature of this idea is that rays never cross bond-surfaces obliquely. The camera shown (*IZ*) has the smallest clearances.

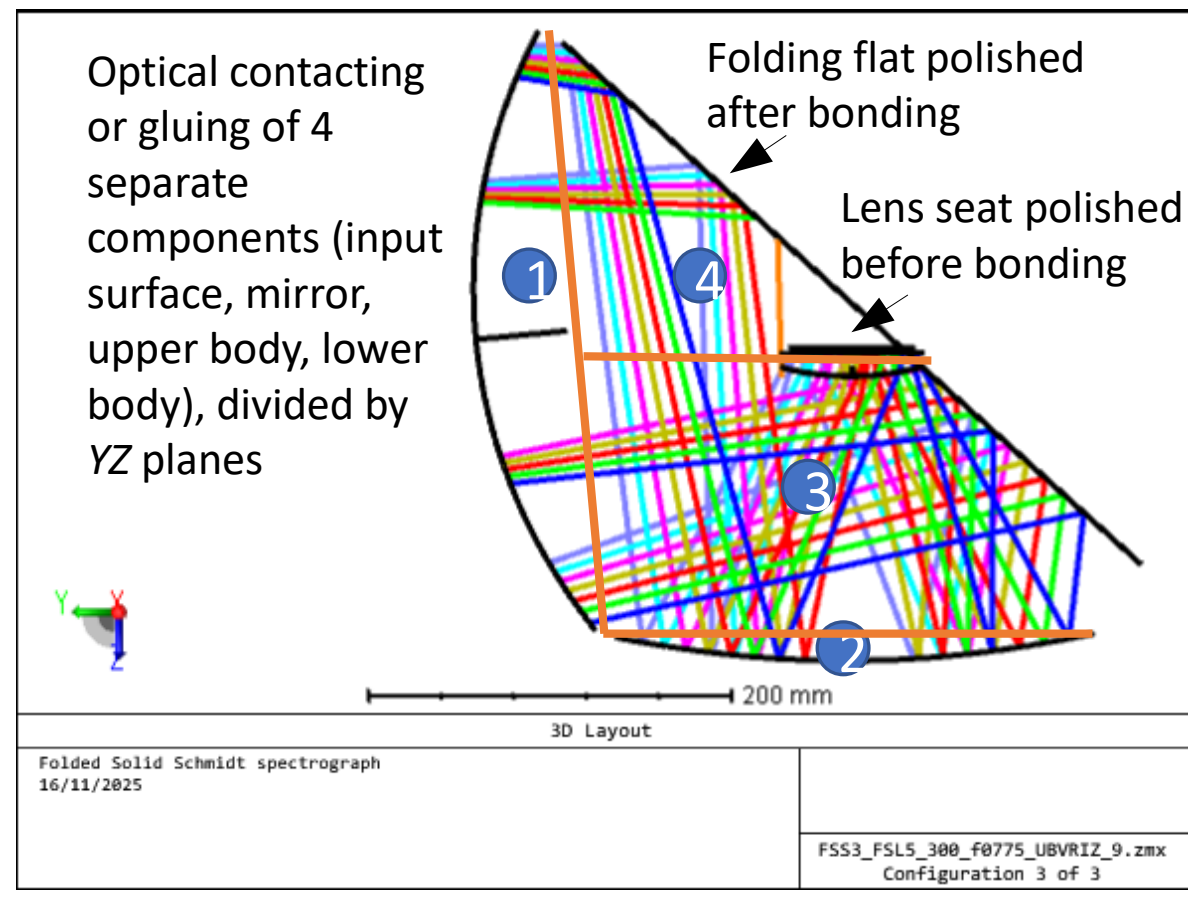

**Figure 7: Proposed method of polishing and assembling the FSS camera body, in the event that in situ polishing of the recessed field-lens seat is not practicable.**

The design was presented to Bertin-Winlight in December 2025. However, due to the novel manufacturing challenges, and the option of lower risk and surely cheaper designs as described below, they declined to quote.

## 5. 4X6 AND 3X9 DIOPTRIC DESIGNS

Dioptric cameras normally offer better throughput than Schmidts, because the detector is no longer in the beam. However, to get few-pixel sampling of the fiber image (as originally deemed necessary) requires camera speeds ~F/1. The fastest transmissive fiber-fed cameras known to us are those for Hector [11], at F/1.31; they have excellent image quality and uniformity, but also many lenses, and the speed is still not fast enough as was desired.

An alternative solution was inspired by Bernard Delabre's first designs for SpecTel [6]. Although having unrealistic detector curvature, they showed the remarkable achromaticity of simple S-FSL5/PBL35Y doublets. Fast flat detector designs based on these doublets have been published for Hector and MSE [12], though neither was adopted, the principal reasons being the large asphericities and tight tolerances, and also the poor options for field-flattener glasses denser than fused silica[4]. A real breakthrough was the availability of large YAG blanks, which allows higher-speed designs with reduced asphericities and increased tolerances.

Fast dioptric designs for WST were initially developed in two variants. The first (4x6) has 4 arms with 6cm detectors, 175mm beam-size and F/1.16 cameras. The second (3x9) has 3 arms with 9cm detectors, 240mm beam-size and F/1.3 cameras. Both have off-axis Schmidt F/3.25 collimators. The cameras have two doublets (or a doublet and a singlet), with matched CTEs. The YAG field flattener is also the dewar window. Although these designs are no longer relevant for WST, they are presented for offering unparalleled speed and simplicity.

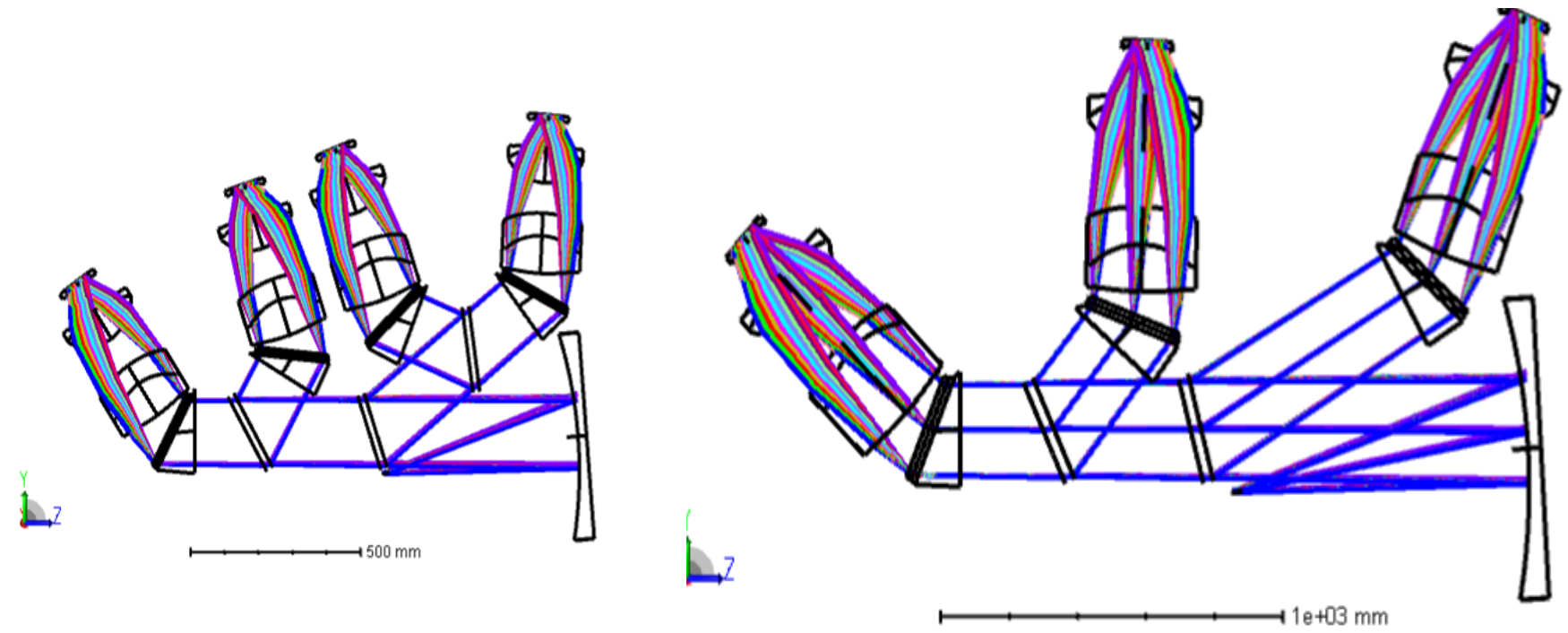

**Figure 8: Layouts for 4x6 and 3x9 designs, to scale, showing the enormous difference in volume and mass.**

The off-axis Schmidt correctors have aperture diameters of 230mm with max aspheric slope 6mrad (4x6), or 300mm with max aspheric slope 7mrad (3x9). They form the input faces of one-sided fused silica grisms incorporating the disperser, This saves two air-glass surfaces and naturally gives non-Littrow use of the disperser, hence eliminating Littrow ghosts. The input and output grating angles are unconstrained, subject to resolution and non-Littrow requirements. The Zemax-obtained solution is always to have $\alpha_1 \approx \beta_0$ in the notation of Baldry et al 2004 [14]. This means there is no anamorphism, so (if the pupil is on the grating) the camera lenses are fully utilised.

The camera barrel is displaced by up to 10mm (4x6) or 20mm (3x9) spatially wrt the beam. It contains two coaxial doublets, with an external asphere on each component. The strongest aspheres are 1.5mm BSFD with slope 140mrad (4x6) or 1.3mm BSFD with slope 75mrad (3x9). The lenses have maximum size of 231mm (4x6) or 331mm (3x9) (assuming 3mm radial mounting margin). All doublets have compatible CTEs. The only glasses used are S-FSL5Y, PBL35Y, and S-FTM16. The thick lenses are all S-FSL5Y or PBL35Y, driven by blank availability from Ohara. All blanks for 4x6 are within Ohara's capabilities, but for the 3x9 design, some lenses would require bonded blanks. The external surfaces of the lenses are all aspheres. Efforts to replace external aspheres with internal ones (with much softer tolerances, and as done for AAOmega [5]), were not successful.

**Field flattener and dewar**

The final optical element in the camera is the YAG field-flattener, which also acts as the dewar window.The field-flatterner is mounted centrally on the dewar. The lens is plano-concave, with (4x6) diameter ~120mm, RoC ~120mm on the 1st surface and asphericity 6-24mrad, or (3x9) diameter ~170mm, RoC ~170mm with asphericity 13-20mrad. The thickness at the centers of the lens is set to 3mm (4x6) or 5mm (3x9). 3mm thickness was tested via FEA for fused silica for Hector, and so should be fine here, given YAGs greater strength. A more detailed FEA study is presented separately [9].The design is tolerant to increasing these thicknesses.

---

4 Granularity and opacity (ALON); birefringence (sapphire); radioactivity (S-YGH51); high dispersion (S-NBH53). Other glasses were tested in [13], none were very good.

The detectors are offset in the spectral direction within the dewar, by up to 5mm (4x6) or 9mm (3x9).

The optimal vacuum gap to the detector is small, just 1mm for 4x6 or 1-1.3mm for 3x9. The design is very intolerant to increasing this gap, but it is difficult to accommodate, because of the requirement for a heat shield. A nice solution is to mill a border mounting strip (say 10mm wide) round the flat lens surface to a depth of a few mm. This leaves a square central 'plug' to the full depth, while the milled border is used for mounting the lens onto the dewar. The O-ring sits in a trough, and must be sized so that the lens sits in full contact with the mounting surface when under vacuum load. This design also ensures correct alignment of the lens on the dewar.

The entire dewar assembly is displaced and tilted spatially wrt to the camera barrel, for focus and alignment.

## Image Quality

The image quality is adequate for both designs, rms radius < (projected fiber diameter)/6 almost everywhere (Figure 9). This is a similar IQ (as a fraction of fiber size) to that for DESI [15], and 'spectroperfectionism [16] or similar would allow precision sky subtraction, along with calibrations which closely mimic the pupil illumination. The image quality is best in the reddest arm, which is helpful both for sky subtraction, and also for minimising pixels lost to *OH* contamination.

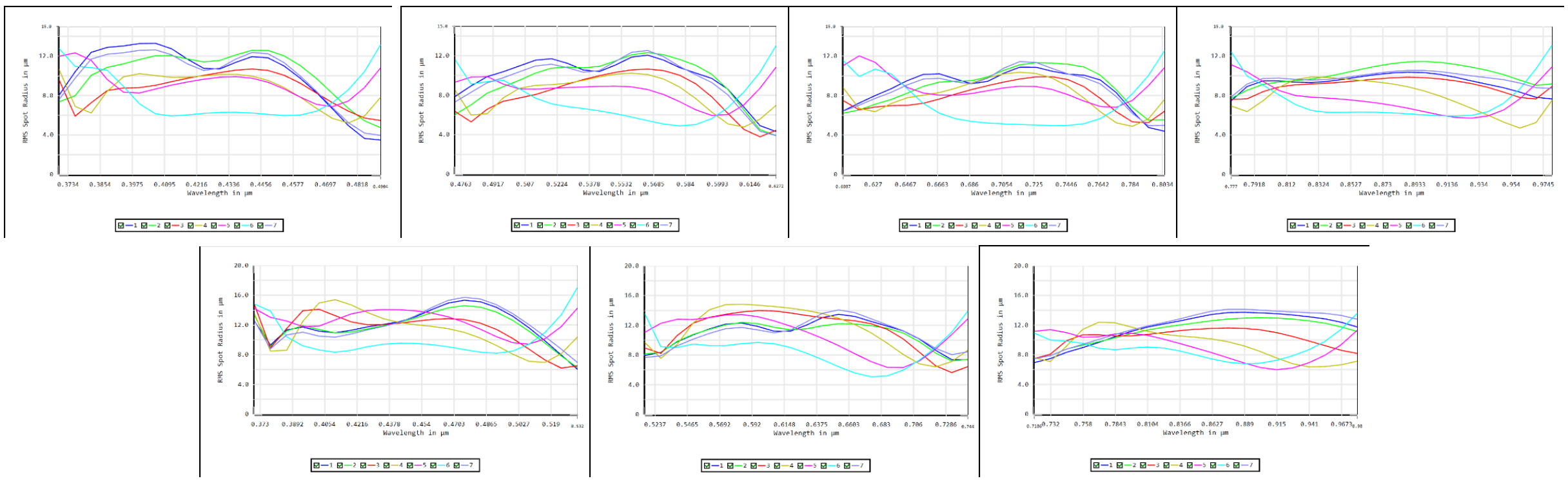

**Figure 9: Rms radius for (above) the 4x6 and (below) the 3x9 designs. Projected fiber size is 71um (4x6) or 78um (3x9).**

## Resolution and throughput

For the 4x6 design, the resolution just meets the $R > 3000$ requirement, while giving a resolution of $R = 4000$ at 850nm. For the 3x9 design, both the minimum resolution and the resolution at 850nm are 5-10% better. Both these design predated the change in wavelength coverage requirement, from 370nm-980nm down to 370nm-930nm, which would increase resolutions by ~5%, and slightly improve the VPH efficiency and image quality also.

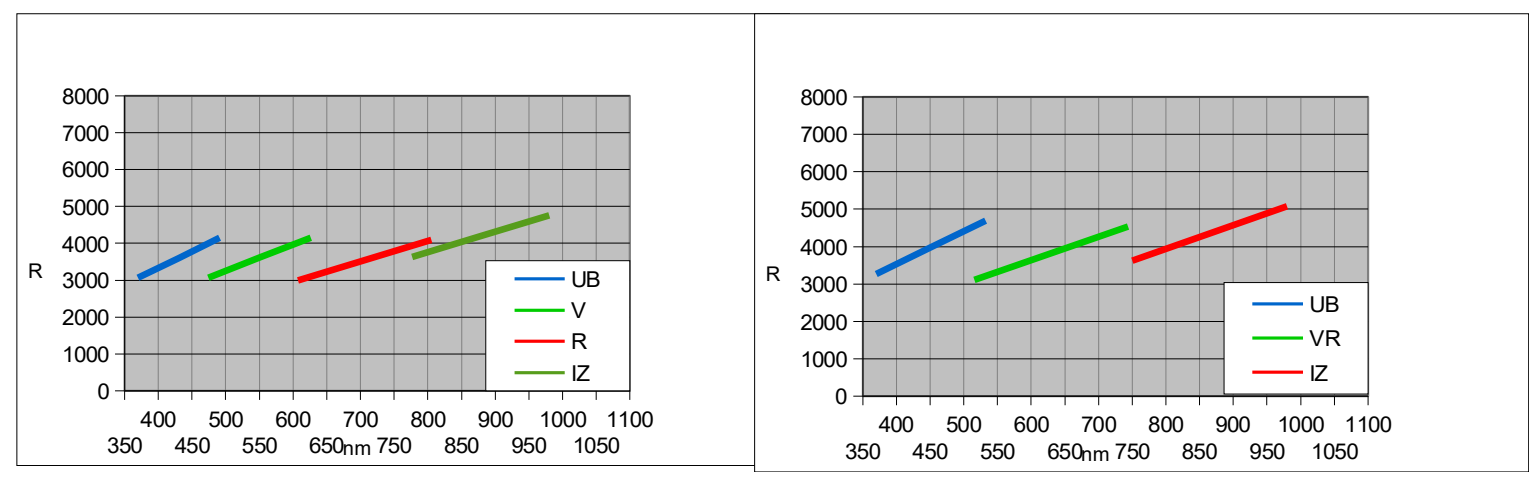

**Figure 10: Resolution for (a) 4x6 and (b) 3x9 designs**

The throughput has not been calculated in detail, but is clearly excellent for both designs. The losses due to glass absorption are just 3.2% (4x6) or 5.5% (3x9) at 370nm. To this must be added the collimator losses, disperser, detector, one mirror, two dichroics and 9 other air/glass surfaces.4x9 dioptric designs.

**Cost and risk**

The designs were presented to Bertin-Winlight in December 2026. The 4x6 design was costed at ~3ME per spectrograph, not including dispersers, detectors or dewars, but also not including the economies of scale for the large numbers of units. So a total unit cost <4M$ is anticipated. The 3x9 was ~10% more expensive overall, but also had higher risk (larger lenses, many of them requiring compound blanks). Since the volume and mass are also larger, and the throughput somewhat worse, the 3x9 design was not pursued further.

## 6. 4X9 DESIGN

**4x9 introduction**

The 4x9 design was initially devised as a lower-risk variant of 4x6, with slower cameras, 9cm detectors, and fused silica field-flattening lenses. The slower speed also gives a small improvement in absolute image quality, and hence a large improvement relative to the projected fiber size. Subsequently, it became clear that 9cm CCD detectors are already not much more expensive than 6cm, and this is expected to be also true for CMOS. Also, predicted CMOS read-noise and dark current are expected to be $<1e^-$/pixel. These two developments essentially removed the 'need for speed' which had been a principal driver for all the earlier designs. At about the same time, WST-IFS selected 9cm CMOS detectors, making this the obvious choice for MOS-LR also.

Two variants were developed, 4x9 and 4x9Y, with the latter reverting to a YAG field lens. At the WST Trade Study in March 2026, these two designs were selected for further development. The location of the LR spectrographs was also partially decided at that time, either above or under the telescope azimuth floor, and rotating with it. This minimises fiber length, but means that precision temperature control is more difficult, both practically and because of the likelihood of impacting the dome seeing. This makes passive temperature control an attractive option, and hence an athermal spectrograph design is strongly preferred. YAG is preferred for an athermal design, both because its CTE (7.9ppm/°C) matches the cameras, and because as a dewar window, it should never suffer from dew or frost (see below). This, and finding an affordable blank supplier (LaserTec), made YAG the preferred field flattener material, with fused silica always available as a backup.

Other design changes have been made:

- The beam size has been increased from 175mm to 184mm, and the maximum lens size to 248mm, the maximum compatible with Ohara blank capabilities;

- The wavelength range has been restricted to 370-930nm.

- The camera speed is now F/1.64, partly driven by a desired slit-length of 175mm, in turn driven by modularity and cross-talk requirements.

- Detector clearance has been increased to 3mm.

Other details are mostly as for the 4x6 design above: the collimator is off-axis, with the 4 cameras arranged on one side, opposite the slit; there are 3 dichroics, the correctors are bonded to a grism containing the disperser; the cameras consist of two bonded doublets and a field-flattening lens which also acts as the dewar window.

The camera doublets have matched CTEs, and are coaxial, for easier alignment. There is an asphere on each lens (5 in total in each camera). The doublet glasses are S-FSL5Y/PBL35Y, or S-FSL5Y/S-FTM16 in the redder cameras. Maximum asphericity is 1.4mm BFSD, 100mrad max slope, and this could likely be reduced to ~1mm with further work.

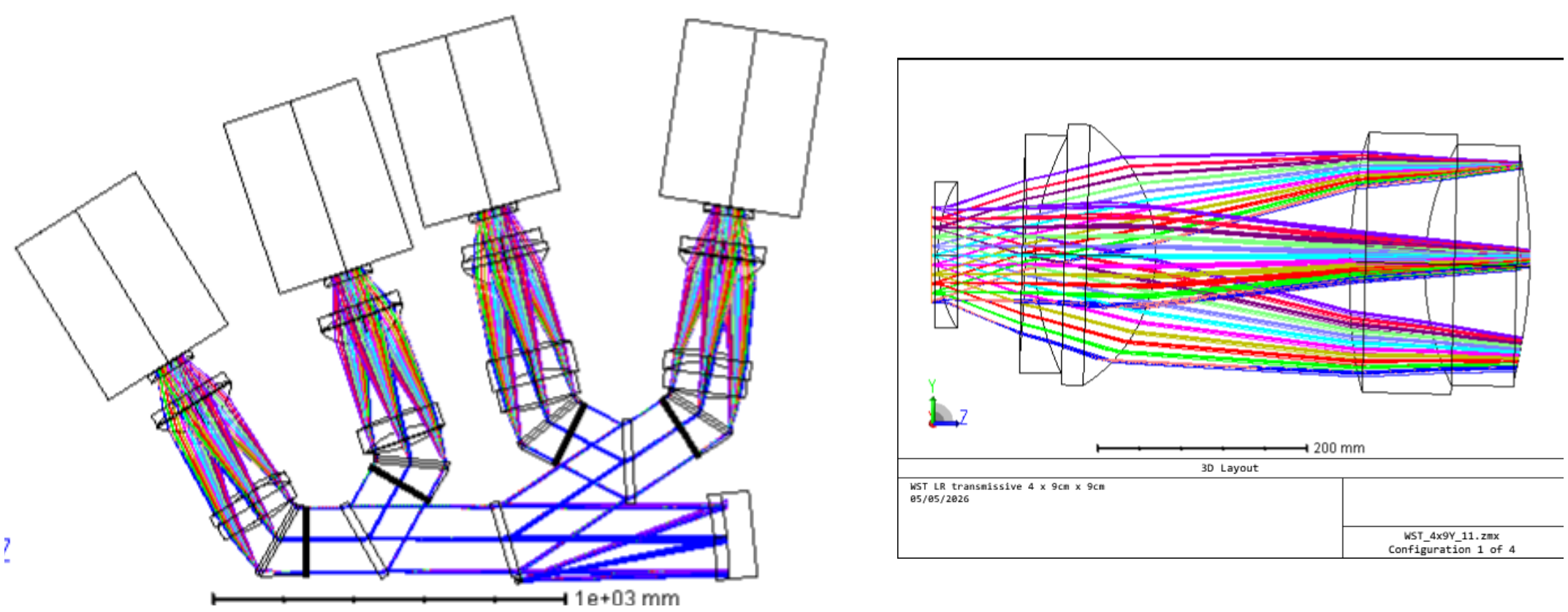

**Figure 11: (a) Layout for the adopted 4x9Y design. The boxes represent nominal dewar volumes. (b) close up of the bluest camera. Doublets are S-FSL5Y/PBL35, (S-FSL5Y/S-FTM16 in the redder cameras). Field flattening lens is YAG. Largest lens diameter is 248mm.**

There is an offset between disperser and camera, an offset and tilt between camera barrel and dewar (adjustable for image alignment and focus), and a small lateral offset of the detector within the dewar.

### Image quality, throughput, resolution, etendue

The 4x9 image quality is better than 8um rms radius almost everywhere (Figure 13), significantly better than the 4x6 and 3x9 designs even in absolute terms, and a dramatic improvement as a fraction of the projected fiber diameter *d* (=100.0um)*,* with rms radius < *d* /12 at all wavelengths and slit positions. It is also significantly better than that for DESI [15] or 4MOST LRS [17] in absolute terms, despite their smaller field angles and slower camera speeds. Evidently, this excellent image quality opens the possibility of reducing the beam and lens sizes to save on cost. But as argued below, an excellent image quality is likely to be necessary for many WST science cases.

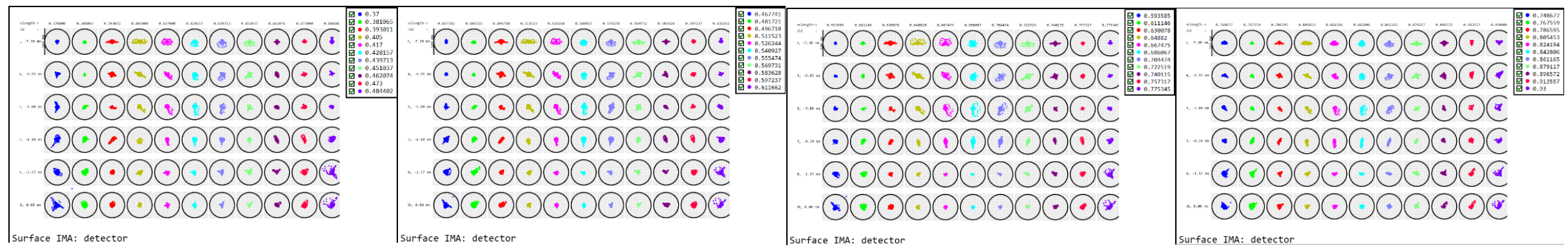

**Figure 12 Spot diagrams for the 4x9 design for the 4 channels. Circle size is the projected fiber size, 100um**

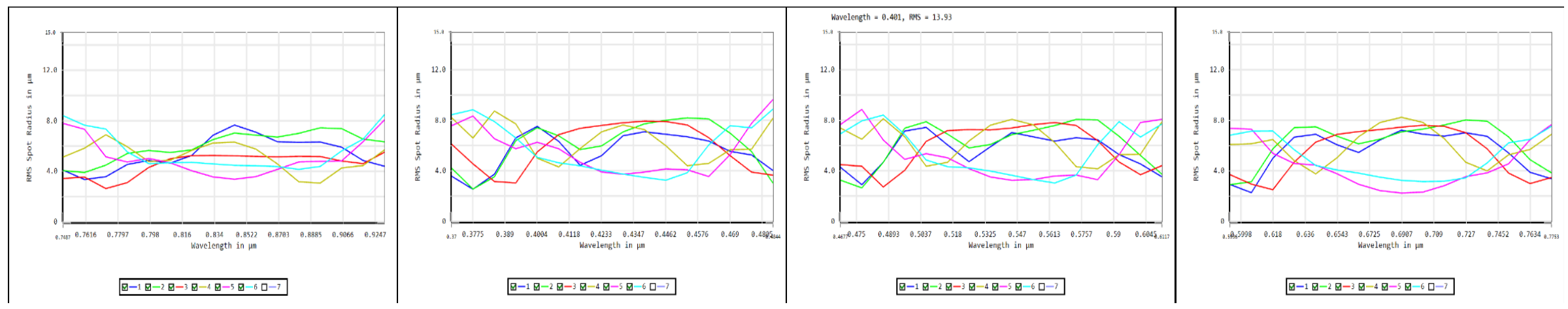

**Figure 13 rms radius vs wavelength for the 4x9 design for the 4 channels.**

The throughput is similar to the 4x6 design, the main losses are due to collimator overfilling (~10%), one mirror and two dichroics, 9 other air/glass surfaces, 200mm of *i*-line glass (3.7% losses at 370nm), plus disperser and detector efficiencies. VPH efficiency is significantly improved (smaller bandwidths and smaller grating angles).

The etendue per spectrograph is 12.5% better than for 4x6, or 3.4 DESIspects. It is also larger than for PFS, and equal even to that of MOONS (and much better if throughout is included).

**Frosting**

For large cryostat windows, dew and frosting are serious concerns. The window is in balance between radiative cooling to the detector, convective heating from the air, and conductive heating from the dewar. The low thermal conductivity and high thermal emissivity of fused silica lead to large gradients. YAG is 9 times more conductive, but the lens is thinner. It is straighforward to directly integrate for the heat transfer, assuming a circular window and neglecting axial temperature gradients. This was done for both fused silica and YAG windows for the 4x9 designs, with results shown in Figure 14(a), and confirmed by thermal FEA analysis (Figure 14b). The net temperature decrement is ~10°C for fused silica, but ~2°C for YAG. Crucially, this is less than the dew point decrement for 80% humidity, at which the dome is closed at Paranal. So as long as the wiindow is in good thermal contact with the dewar, dew and frost should in principal never arise.

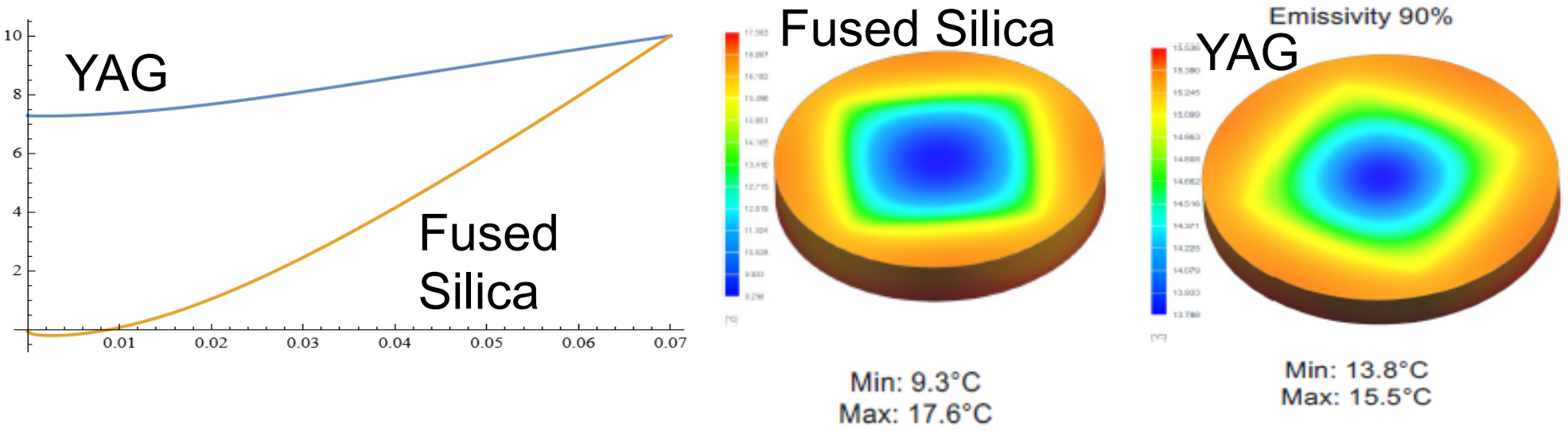

**Figure 14: Temperature decrement for fused silica and YAG windows of the shapes required for 4x9; (a) by direct integration assuming a circular window and no axial gradients, and (b) a full FEA thermal analysis assuming square windows.**

**Thermal stability**

It happens that the CTE of all camera lenses is close to 8ppm/°C, and this is not so far from the CTE of ferritic stainless steel, or standard optical benches, or float glass (for the collimator mirror), all of which are close to 10ppm/°C. The dispersers are on fused silica, with CTE just 0.5ppm/°C, so thermal changes in the grating density are small. So the 4x9 design is intrinsically insensitive to temperature changes. It was then natural to see if a truly athermal design could be devised: Figure 15 shows the rms radius without refocus for the bluest arm (the most difficult) at -8°C, 8.5°C and 25°C, representing the absolute temperature range at Paranal. It practice, the expected temperature variations in the insulated spectrograph chamber are only a third as large, say 5-15ºC, and changes in image quality should in principal be almost undetectable.

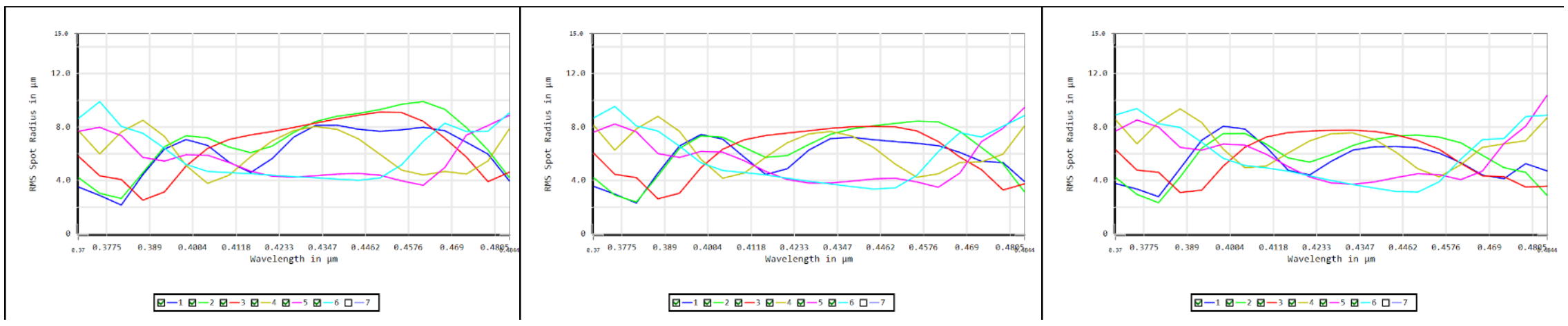

**Figure 15: Rms radius for Paranal temperature extremes of -8°C,8.5°C,25°C**

**Cost and risk**

The design is currently with Bertin-Winlight for an industrial study (RoM costs, timescales, required equipment, facilities, staff, etc.) A similar design proposed for WST MOS-HR has been costed at 3ME, so it seems that the 4x9Y design is not more expensive than 4x6, and the same 4ME/unit cost represents a conservative estimate.

## 7. FSF STABILITY

Accurate knowledge of the FSF (the image of the fiber at the detector) is essential for many reasons (sky subtraction, wavelength calibration, cross-talk correction, detailed LSF analysis), and the larger the fiber etendue, the more extreme the requirements. For 12m telescopes in natural seeing, an precision of <<1% is desirable [18]. WST has an additional challenge that as a facility telescope combining many different science programs and target lists, large magnitude differences are inevitable, with consequent cross-talk issues; also FSF requirements are driven by the most demanding science programs. This level of FSF precision looks extremely challenging, requiring both excellent mimicry of the telescope pupil illumination between science and calibration observations, and also excellent spectrograph IQ, which reduces the sensitivity of the FSF to pupil illumination errors.

A simple proxy for this sensitivity can be determined by comparing the FSF with and without apodisation of the fiber far-field. Figure 16 shows the difference between the FSF for apodised and unapodised fiber far-field output, scaled to the peak FSF height; differences of several per cent are evident. While this greatly exaggerates the likely imperfections of the calibration system, it demonstrates that achieving <<1% precision on knowledge of the FSF shape will require both pupil illumination by the calibration system closely mimicking the telescope (say to within a few %), and an exceptional spectrograph IQ.

**Figure 16: Difference between the Fiber Spread Function for the 4x9 design (blue end and end-of-slit) for apodised and unapodised fiber far-field input to the spectrograph, scaled to the peak FSF height. The maximum difference exceeds 4%, even for this design with excellent IQ.**

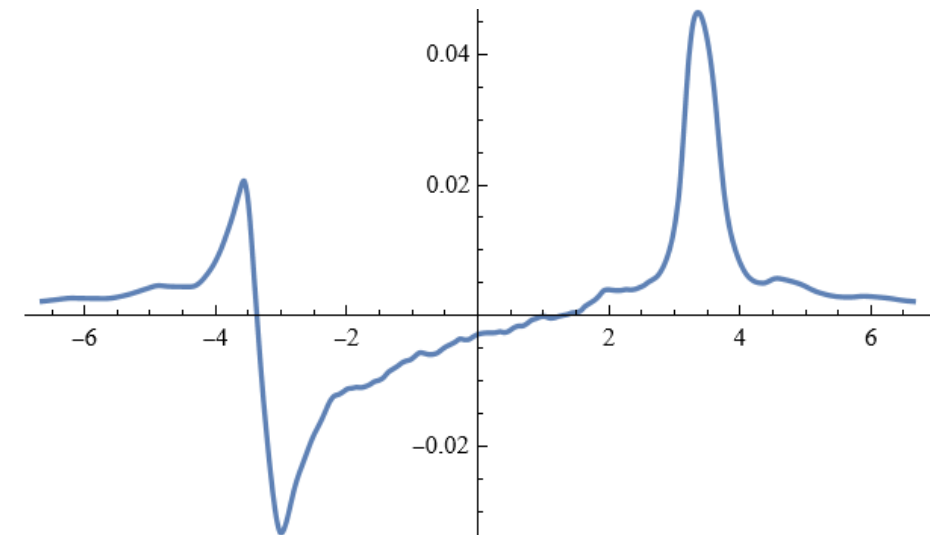

## 8. SCATTERING, FWHM, CROSS-TALK AND FIBER NUMBERS

The MOS LR spectrographs form a large component of the total cost of WST, so we must put as many fibers into each spectrograph as efficient spectral extraction allows. The pitch is primarily limited by the cross-talk between adjacent fibers, which diminishes the S/N of both spectra (even assuming that the FSF is perfectly known). This cross-talk results from the spectrograph aberrations, charge diffusion, detector flatness., and scattered light We have determined the minimum pitch on the basis that the measured cross-talk performance of 4MOST LRS should form the requirement for WST.

Assuming equally bright targets, perfectly known FSF and no read-noise, then the average S/N hit due to pixellation, cross-talk and scattering from all other fibers, averaged over all possible pixellations, is simply given by

$$S_T = \int f'(x)^2 / \Sigma_i f'(x + i\,p) \; dx \tag{1}$$

where $f'$ is the spatial FSF, boxcar-smoothed by one pixel; $x$ is position in pixels, $p$ is the pitch in pixels, $i$ is the relative fiber number, and the sum is over all fibers. The dependence on the precise pixellation is very weak, with typical variation < 1%, and can be ignored.

This metric has various advantages:
- it closely reflects what the extraction software does.
- it gives the observing efficiency penalty directly, very helpful for e.g. the Exposure Time Calculator.
- it allows straightforward trade-off analysis between pitch, cross-talk, survey speed and fiber numbers.
- is easily calculated for both modelled and actual data.
- by restricting $i$ to {-1,0,1}, one obtains $S_x$, the cross-talk just due to adjacent fibers (which dominates systematic spectral error).

The actual spec can be phrased along the lines of: 'The loss of observing speed due to cross-talk, scattered light and detector pixellation, assuming constant and equal flux per unit wavelength and no read-noise, shall not exceed 10% (5% goal) compared with a spectrograph with perfect optics and no detector pixellation, over 95% (goal 99%) of each detector.'

Figure 17 shows a by-eye model fit to a spatial profile for 4MOST LRS-A Blue data for a bright star. This seems to do a reasonable job at reproducing the observed profile, both near and far. The fit components are the projected disc of the fiber, convolved with (a) a one-pixel boxcar, (b) a gaussian representing aberrations (both as-designed and as-built), and also charge diffusion and detector non-flatness, (c) $1/x^3$ PSF wings, (d) $1/x$ scattered light; there is also (e) a fixed scattered light component. A Lorentzian $1/x^2$ term was originally included but seems redundant. Assuming that the spectral profile is similar, the 'spectral purity' (fraction of the light falling within twice the FWHM) is 94%.

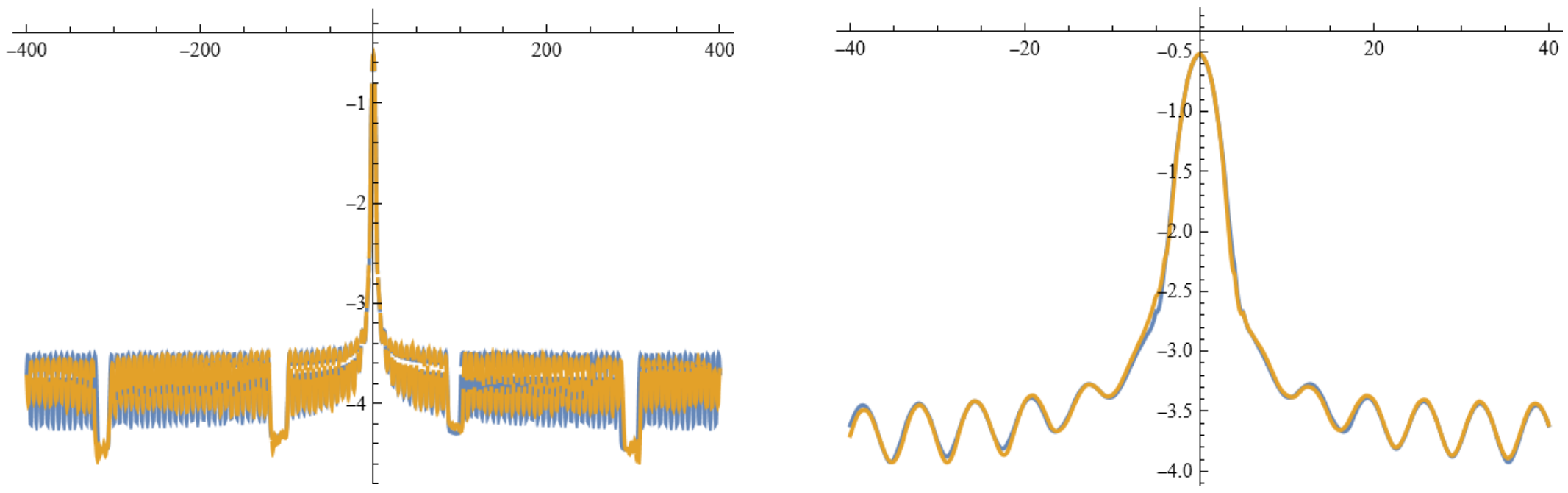

**Figure 17: (a) Co-added spatial profile through 4MOST LRS blue channel data, with a bright star in one fiber. The data is orange, while the model fit is in blue. The imperfect fit at large distances is due to the changing FSF of the spectrograph. (b) A close-up of the central fiber.**

The scattered light profile derived for 4MOST LRS was then applied to each of the designs presented here, modified by the differences in projected fiber diameter, camera focal length and rms radius. This allows a minimum pitch to be derived, assuming the same level of cross-talk between fibers as achieved for 4MOST LRS ($S_T = 0.9$). Note that while the pitch-to-fiber-diameter ratio is smaller than for existing MOS systems on smaller telescopes, the physical gap between fibers is comparable or larger, and it is essentially this gap that determines the cross-talk.

Knowing the minimum pitch then allows the maximum number of fibers per spectrograph to be derived, and also the minimum number of spectrographs to accommodate 30K fibers (Table 1). These numbers assume that 5% of the slit is lost to gaps imposed by the slit and slitlet geometry, or for measuring scattered light, or for calibration fibers.

**Table 1. Pitch, fiber and spectrograph numbers for the various designs, assuming 195um fibers**

| | Camera focal length (mm) | Projected fiber diameter (um) | IQ (rms radius, um) | Pitch/ diameter ($S_T = 0.9$) | Slit length (mm) | NFibers per spect | NSpect (30K fibers) |
|---|---|---|---|---|---|---|---|
| FSS | 232 | 49 | 6 | 1.65 | 245 | 725 | 41 |
| 4x6 | 201 | 70.5 | 12 | 1.56 | 164 | 513 | 58 |
| 3x9 | 311 | 79.5 | 14 | 1.74 | 225 | 628 | 48 |
| 4x9Y | 300 | 100 | 8 | 1.51 | 175 | 565 | 54 |

In practice, modularity and cost means that there are likely to be only 28K MOS-LR fibers in total, and 525 fibers per 4x9Y spectrograph, leading to 54 spectrographs with a pitch/diameter ratio ~1.63, giving an expected cross-talk performance significantly better than 4MOST LRS. All these calculations depend on the optics having the same scattered light profiles as 4MOST. If surface relief gratings offer improved scattered light performance, as seems likely, then the cross-talk could be dramatically improved, and/or fiber numbers per spectrograph increased and total costs reduced.

## ACKNOWLEDGEMENTS

We are extremely grateful to the following people for their invaluable help in developing these designs. Jonathan Irwin at IoA Cambridge provided the 4MOST spatial profiles that were used to analyse the scattered light and derive cross-talk estimates for WST. Roland Winkler of AIP and Dioone Haynes at ANU provided 4MOST FRD images. Richard McDermid provided the Hanushik sky spectrum in readable form. Gary Liu of LaserTec provided a timely and affordable quote for YAG blanks in the required sizes. Elroy Pearson of Wasatch Photonics provided the VPH efficiency curves in Figure 1; Martin Rumpel (IoF) modelled SR grating efficiencies and scattering for us; Chris Ghio of Ohara US obtained maximum feasible blank sizes for all the relevant glasses (which were instrumental in developing these designs); all three patiently answered endless questions. Vincent Lapere and Julien Marque of Bertin Technologies have also given invaluable feedback.

## APPENDIX - FAR RED RESOLUTION AND OH CONTAMINATION

The WST MOS-LR resolution requirement is $R > 3000$ at all wavelengths. But a higher resolution would clearly be desirable in the far red (say beyond 860nm) to work between *OH* lines, and also to provide 1km/s stellar radial velocities from the calcium triplet. To test the effect of resolution on survey efficiency, the spatial FSF model derived from 4MOST LRS was adapted to the 4x6 design, and convolved with the UVES sky spectrum [], for various nominal resolutions. The inverse of the resulting flux makes an excellent proxy for survey speed for spectral features at that wavelength, and is shown in Figure 18. Gains of about 5% accrue in going from 2.5nm to 1.75nm resolution (Table 2).

Surface relief gratings offer the possibility of very significant reduction in scattered light (normally dominated by the disperser), so as a thought-experiment, also shown is the sensitivity for the case of no scattering. Much more dramatic gains are possible, about 13% at all resolutions. This is in addition to any direct efficiency gains offered by the surface relief gratings.

**Table 2. Total survey speed, integrated across the *Z*-band, relative to the speed at the UVES resolution**

| Resolution (nm) | 0.175 | 0.2 | 0.225 | 0.25 |
|---|---|---|---|---|
| Resolution *R* @860nm | 4910 | 4300 | 3822 | 3340 |
| *Z*-speed VPH | 0.741 | 0.728 | 0.715 | 0.703 |
| *Z*-speed no scatt | 0.840 | 0.823 | 0.808 | 0.792 |

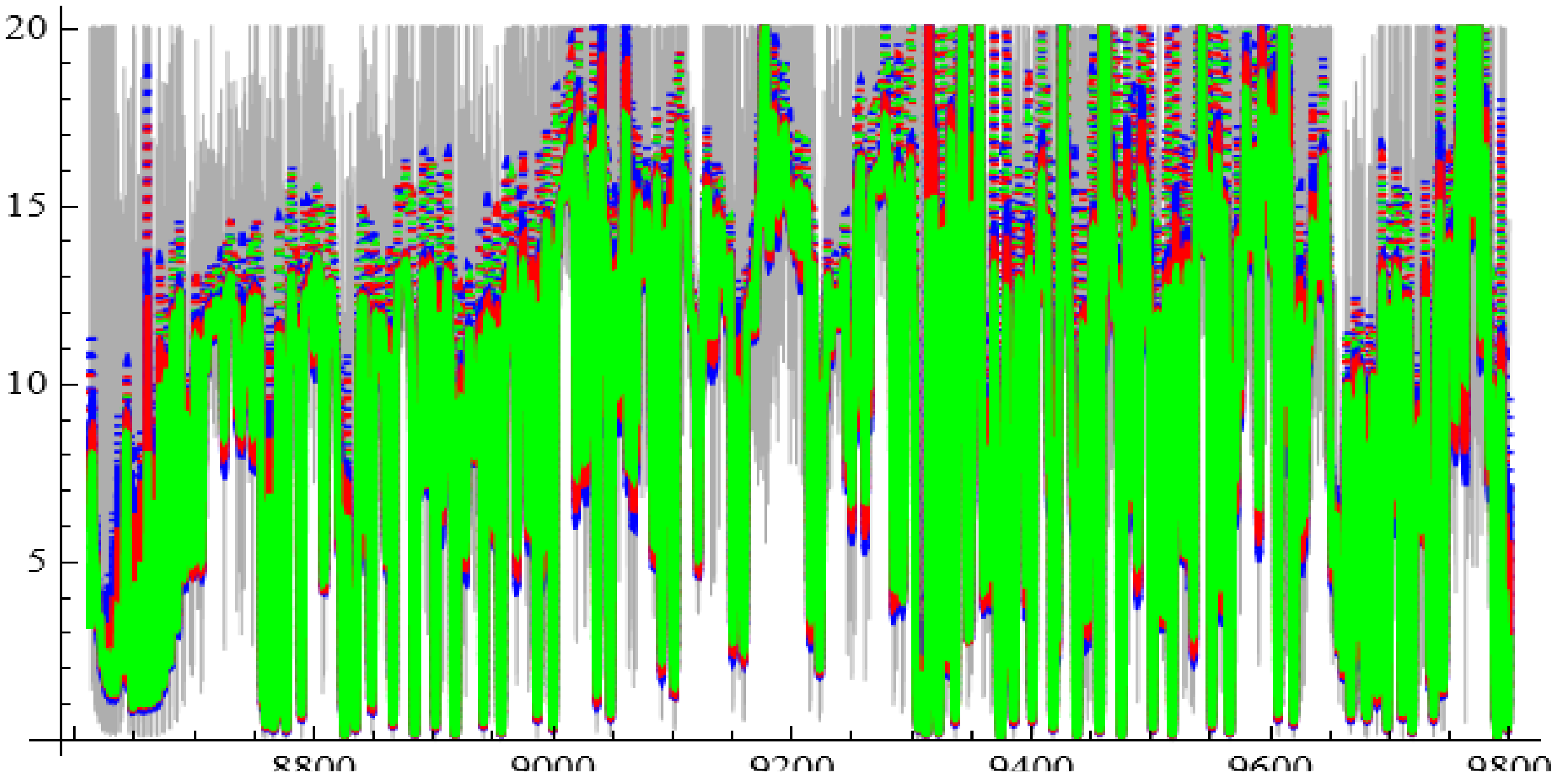

**Figure 18. Inverse sky flux vs wavelength, from the UVES spectrum of Hanuschik 2003 [19] (grey), and when convolved with a PSF derived from 4MOST LRS for 2.5nm, 2nm and 1.75nm FWHM resolution (green/red/blue solid lines). Also shown are the results with no scattering (dotted lines).**